# Characterization and quality control test of a gigabit cable receiver ASIC (GBCR2) for the ATLAS Inner Tracker Detector upgrade

**W. Zhang,**[a,b,1] **C. Chen,**[a,b,1] **D. Gong,**[b] **S. Hou,**[c] **G. Huang,**[a] **X. Huang,**[a,b,1] **C. Liu,**[b] **T. Liu,**[b,*] **H. Sun,**[a,b,1] **X. Sun,**[a] **P. Wang,**[b] **J. Ye,**[b] **and L. Zhang**[a,b,1]

[a] *Central China Normal University,*
  *Wuhan, Hubei 430079, P.R. China*

[b] *Southern Methodist University,*
  *Dallas, TX 75275, U.S.A.*

[c] *Institute of Physics,*
  *Academia Sinica, Taiwan.*

  *E-mail:* tliu@smu.edu

ABSTRACT: We present the characterization and quality control test of a gigabit cable receiver ASIC prototype, GBCR2, for the ATLAS Inner Tracker pixel detector upgrade. GBCR2 equalizes and retimes the uplink electrical signals from RD53B through a 6 m Twinax AWG34 cable to lpGBT. GBCR2 also pre-emphasizes downlink command signals through the same electrical connection from lpGBT to RD53B. GBCR2 has seven uplink channels each at 1.28 Gbps and two downlink channels each at 160 Mbps. The prototype is fabricated in a 65 nm CMOS process. The characterization of GBCR2 has been demonstrated that the total jitter of the output signal is 129.1 ps (peak-peak) in the non-retiming mode or 79.3 ps (peak-peak) in the retiming mode for the uplink channel and meets the requirements of lpGBT. The total power consumption of all uplink channels is 87.0 mW in the non-retiming mode and 101.4 mW in the retiming mode, below the specification of 174 mW. The two downlink channels consume less than 53 mW. A quality control test procedure is proposed and 169 prototype chips are tested. The yield is about 97.0%.

---

[1] Visiting scholar at SMU and performed this work at SMU.

[*] Corresponding author.

# Contents

## 1. Introduction

The High Luminosity Large Hadron Collider (HL-LHC) is expected to begin operation in 2026 [1] with a design instantaneous luminosity of $5\times10^{34}$ $cm^{-2}s^{-1}$. The ATLAS detector, including Inner Tracker (ITk) [2], will be upgraded (Phase-II) [3, 4] accordingly. In the ATLAS ITk pixel upgrade, each pixel module has a pixel sensor and two or four RD53B chips [5, 6, 7] in parallel. Each readout chip RD53B transmits the detector data at 1.28 Gbps per channel to a low-power Gigabit Transceiver (lpGBT) [8]. The lpGBT congregates the data to 10.24 Gbps per channel. An optical module Versatile Transceiver (VTRx+) [9] converts electrical signals to optical signals and then sends optical signals to the counting room through 80 m optical fibers.

The HL-LHC upgrade leads to a significant radiation-level increase in the ITk pixel detector. As a result, the VTRx+ and the lpGBT must be placed a few meters away from the detector. As the pixel detector is located innermost in the ATLAS detector, it is critical to minimize its mass budget. A section of flexible printed circuit from 0.1 m to 1 m followed by an American Wire Gauge (AWG) 34 Twinax cable of up to 6 meters is chosen to connect the RD53B with the lpGBT. Because the cable's loss at 1 GHz is as high as 25 dB, the RD53B's output signal is completely embedded in the noise after passing through such a thin cable and the lpGBT cannot identify it. A gigabit receiver is needed to recover the signal's high-frequency loss for the lpGBT.

To meet the demand of electrical data transmission through thin cables, two prototype ASICs, GBCR1 [10, 11] and GBCR2 [12] have been designed. The first prototype ASIC called GBCR1 is designed for the initial data transmission scheme, in which each ASIC receives 4 uplink channels each working at 5.12 Gbps. After the test of GBCR1, the transmission scheme has been evolved into the current baseline scheme, in which 6 uplink channels each operating at 1.28 Gbps are handled in each ASIC. The second ASIC prototype named GBCR2 is developed as the current baseline transmission scheme. GBCR2 includes seven uplink channels to match the 1.28 Gbps channel number of the lpGBT.

In addition to the uplink channel, GBCR2 optionally provides two downlink channels. Each downlink channel transmits commands at 160 Mbps from the lpGBT to the RD53B through the same cables as those used in uplink channels. The pre-emphasis of the lpGBT may not be strong enough to drive such a thin cable, so two downlink channels with the pre-emphasis function are integrated into GBCR2 to deliver commands from the lpGBT to the RD53B. Reducing the jitter of downlinks is also helpful to alleviate the jitter of uplinks because the clock of uplinks is recovered from the Clock and Data Recovery (CDR) unit of the RD53B [13]. The ultimate requirement of the data transmission is that the bit rate error (BER) of uplinks is less than $10^{-12}$.

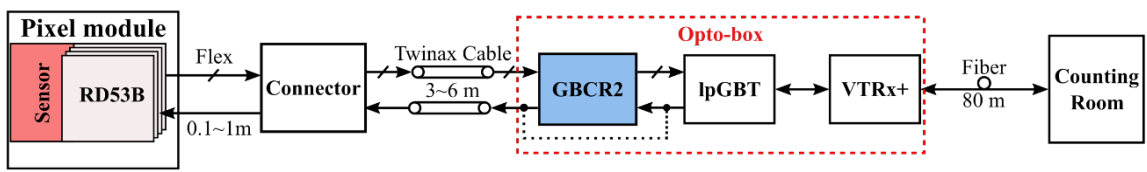

Figure 1. Block diagram of the ATLAS ITk readout system.

The block diagram of the ATLAS ITk pixel readout system is shown in figure 1. GBCR2, as well as lpGBT and VTRx+, are installed in a mini-crate called Opto-Box [14]. In this paper, we present the characterization and the Quality Control (QC) test of GBCR2.

## 2. Design

The block diagram of GBCR2 is shown in figure 2. Each uplink channel equalizes the signal from the Current-Mode Logic (CML) driver of the RD53B through a long cable. The equalized signal can be optionally retimed with a 1.28 GHz phase-adjustable clock provided by a phase shifter. Seven uplink channels share the phase shifter. The equalized or retimed data are driven to the electrical receiver (eRx) [15] of the lpGBT. Each downlink channel pre-emphasizes the signal from an electrical transmitter (eTX) port of the lpGBT and transmits the signal to the eRx port of the RD53B through the same cable as those used in uplink channels. An $I^2C$ target is used to configure the uplink channels, the downlink channels, and the phase shifter.

Each uplink channel consists of a passive attenuator, an equalizer, a limiting amplifier (LA), a DC-offset-cancellation circuit, a retiming logic, and a CML driver. The programmable passive attenuator aims at avoiding the saturation of the following equalizer. The attenuation is adjustable with 1/3, 2/3, and 1. The long cable attenuates high-frequency components more than low-frequency components, resulting in significant Inter-Symbol Interference (ISI) jitter on the receiver side. The equalizer behaves like a high-pass filter, degenerates the low-frequency components, and correspondingly compensates for the high-frequency loss caused by the long cable. The simulated frequency-response curve of the cable in the worst case (1 m Flex cable and 6 m Twinax cable) is shown in figure 3(a). The decay slopes in the low (80 to 200 MHz), medium (200-400 MHz), and high (400 MHz-1.3 GHz) frequency ranges are 8.5, 14.3, and 27.3 dB/dec, respectively. The equalizer of each uplink channel is implemented in Continuous-Time Linear Equalizers (CTLEs). The block diagram of the whole equalizer in an uplink channel is shown in figure 3(b). The whole equalizer is composed of three identical high-frequency CTLEs, a medium-frequency CTLE, and a low-frequency CTLE. The schematic of a single CTLE stage is shown in figure 3(c). Each CTLE is a source-coupled differential amplifier with source degeneration. The high-frequency boosting is achieved by introducing a real zero with a parallel resistor-capacitor ($Rs - Cs$) network. Rs can be tunable to adjust the peaking strength. The three high-frequency CTLEs and the medium-frequency CTLE have adjustable resistance (HFSR and MFSR as shown in figure 3(b)). The whole equalizer provides up to 30 dB peaking strength. The power consumption of the equalizer is about 0.84 mW per channel.

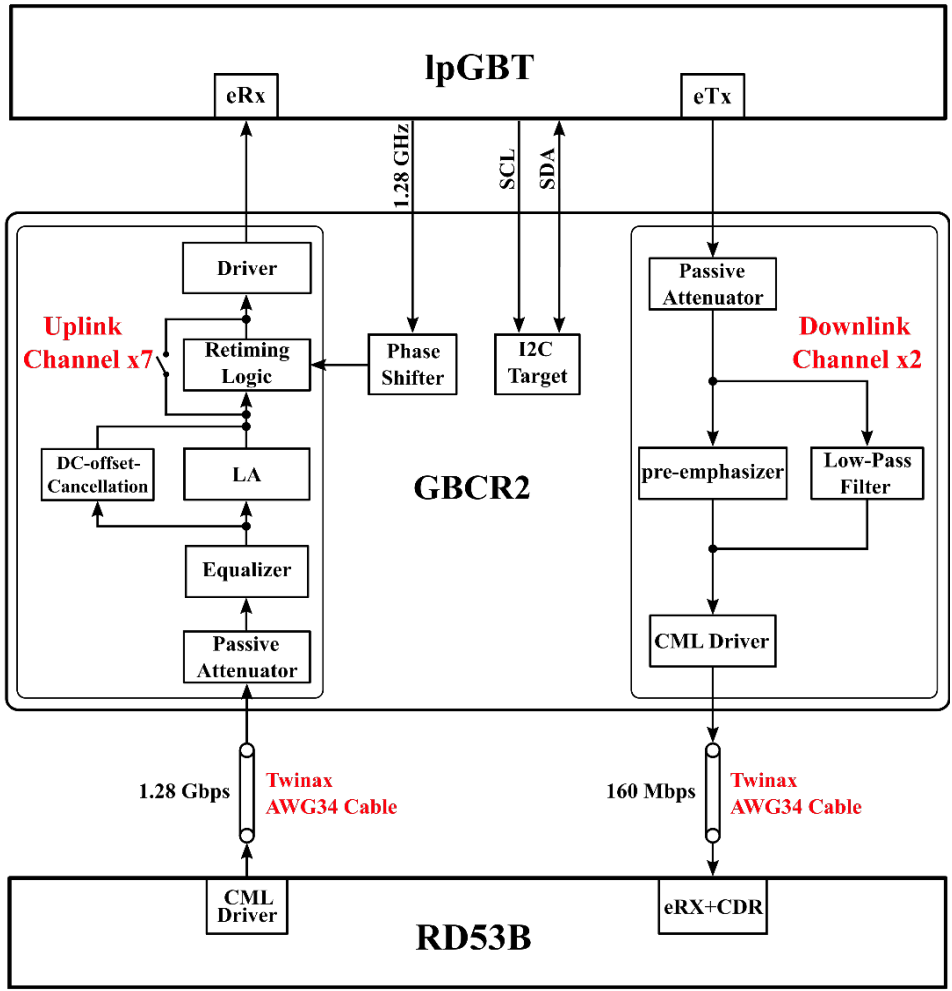

Figure 2. Block diagram of GBCR2 and its interface.

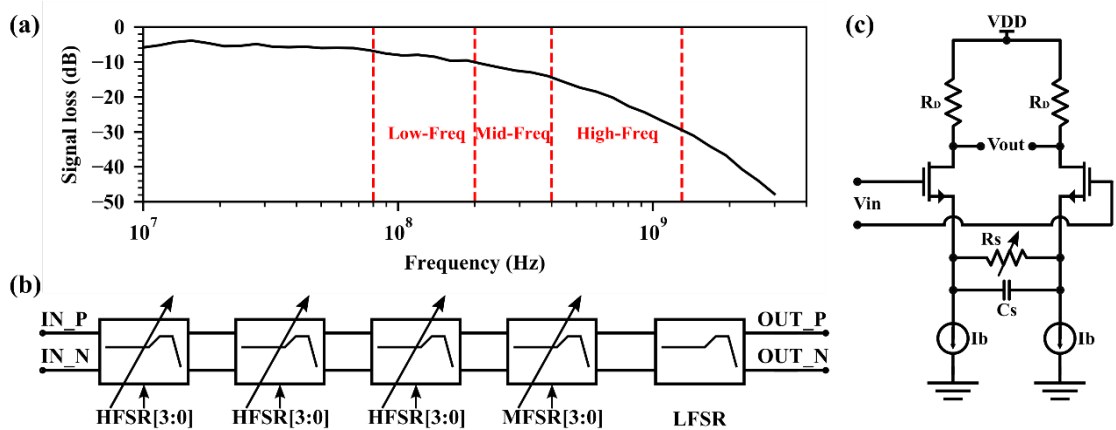

Figure 3. Simulated frequency-response curve of a 1 m Flex and 6 m Twinax cable (a), block diagram of the whole equalizer (b), and Schematic of a CTLE stage (c).

The LA amplifies the equalizer's output signal to saturation for the output driver. The schematic of the LA is shown in the top part of figure 4. A structure with two identical amplification stages is adopted to trade off gain against bandwidth. Each stage is a common-source differential amplifier. The LA includes a DC-offset-cancellation circuit to remove the DC offset caused by the mismatch of the LA and the equalizers. The schematic of the DC-offset-cancellation circuit is shown in the bottom part of figure 4. The high-frequency components of the LA output are filtered by an active low-pass R-C circuit and the DC offset is fed back to the input stage of the LA. The low cut-off frequency of the low-pass R-C circuit is about 115 kHz, low enough for the 1.28 Gbps DC-balanced signal. The operational amplifier (Amp) in the DC-offset-cancellation circuit is used to increase closed-loop stability. The gain and bandwidth of the LA are 7.97 dB and 3.08 GHz, respectively. The total power consumption of the LA is about 0.28 mW per channel.

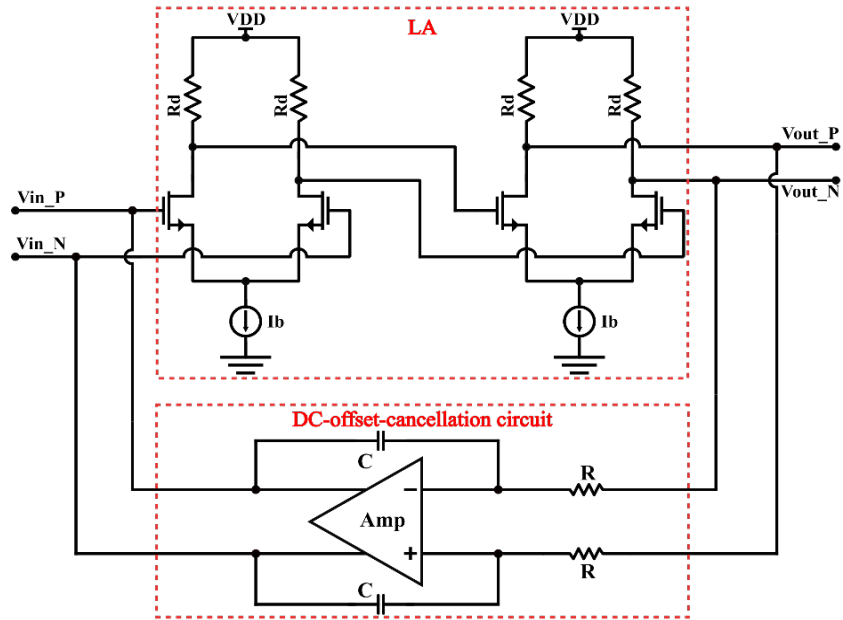

Figure 4. Schematic of the LA, including the DC-offset-cancellation circuit.

The uplink channels have two operational modes, the retiming mode and the non-retiming mode. In the retiming mode, the equalized signal is retimed to reduce further its jitter at the cost of higher power consumption. A phase shifter receives a 1.28 GHz clock from the lpGBT and generates the phase-adjustable clocks for each channel's retiming logic to sample the data. In the non-retiming mode, the retiming logic is bypassed and the phase shifter is disabled.

The block diagram of the phase shifter is shown in figure 5. The core of the phase shifter includes a Delay-Locked Loop (DLL) and eight multiplexers. The DLL consists of a Phase Detector (PD), a Charge Pump (CP), a Low-Pass Filter (LPF), and a Voltage Controlled Delay Line (VCDL). The VCDL divides a 1.28 GHz clock period into 16 equally distributed phases with a resolution of 48.8 ps. Eight multiplexers are used to select different phases for seven uplink channels and a test channel. The phase shifter generates eight phase-adjustable clocks, one for test purposes and the others for seven uplink channels. The phase delay of each uplink channel can be independently configured through the I$^2$C interface. The power consumptions of each channel's retiming logic and the phase shifter are 1.20 mW and 18.36 mW, respectively.

The CML driver is able to transmit the recovered data to the lpGBT. The schematic of the CML driver is shown in figure 6. The CML driver is a differential amplifier with a passive load of 50 Ω. The tail current ($Ibias$) is adjustable to tune the output amplitude from 44.6 mV to 313.7 mV. The power consumption of the CML driver ranges from 2.16 mW to 14.4 mW per channel.

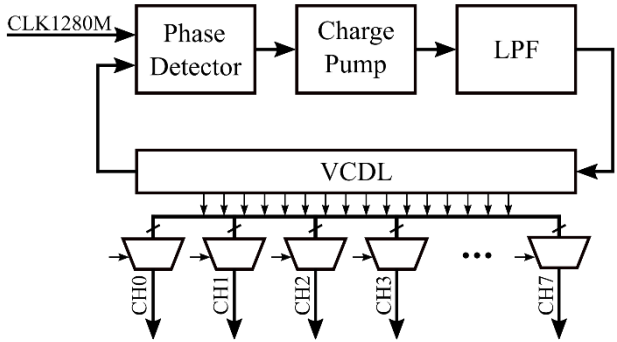 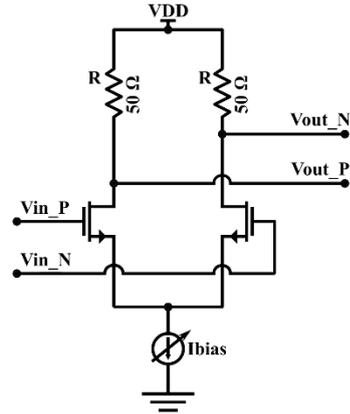

Figure 5. Block diagram of the phase shifter.   Figure 6. Schematic of the CML Driver.

Each downlink channel includes a passive attenuator, a pre-emphasizer, a low-pass filter, and a CML driver. The passive attenuator has three attenuations, 1/3, 2/3, and 1. The pre-emphasis has two identical stages of CTLEs, each with the same structure as the uplink channels. The CTLEs stages are optimized for the data rate of the downlink channel's data rate, which is significantly lower than that of the uplink channels. Each CTLE with a zero at 40 MHz and a pole at 120 MHz. The CTLE stages with an adjustable source resistance provide an emphasis strength up to 14.8 dB. The CTLE stages can be optionally disabled so that the signal passes through without any pre-emphasis. The low-pass filter circuit serves as feedback for the pre-emphasizer to eliminate the DC offset. The output driver with maximum drive strength ($Ibias\_max$) drives the long cable.

GBCR2 is designed in a 65 nm CMOS technology with a single power supply of 1.2 V. The dimension of GBCR2 dies is 1 mm × 2 mm. GBCR2 is packaged in a 48-pin Quad-flat No-leads (QFN) plastic package with a pin pitch of 0.4 mm and a dimension of 6 mm × 6 mm.

## 3. Test results

### 3.1 Test setup

The block diagram of the test setup is shown in figure 7(a). For the uplink channels, a pattern generator (Hewlett Packard Model 8133A) provides a 1.28 Gbps Pseudo-Random Binary Sequence (PRBS) $2^{31}$-1 signal with an amplitude of 200 mV (peak-peak) and a 40 MHz trigger clock. A 6 m AWG34 Twinax cable connects the pattern generator with the test board. A socket is installed on the test board so that we can test different chips. An oscilloscope (Tektronix Model DSA70804B) is utilized to observe eye diagrams and measure jitter. An error detector (Anritsu Model MP1764C) is used to measure the BER. For the downlink channels, the same pattern generator generates a 160 Mbps PRBS signal with an amplitude of 200 mV (peak-peak) to the test board and a 40 MHz trigger clock for the oscilloscope. The Twinax cable is moved to connect the test board with the same oscilloscope. For the uplink and the downlink channels, a laptop running a python script controls the test setup and records the supply current. The Radio Frequency (RF) switch (MUX) and the USB-to-digital-I/O adaptor are used in the QC test and will be discussed in the next section. A photograph of the test setup is shown in figure 7(b).

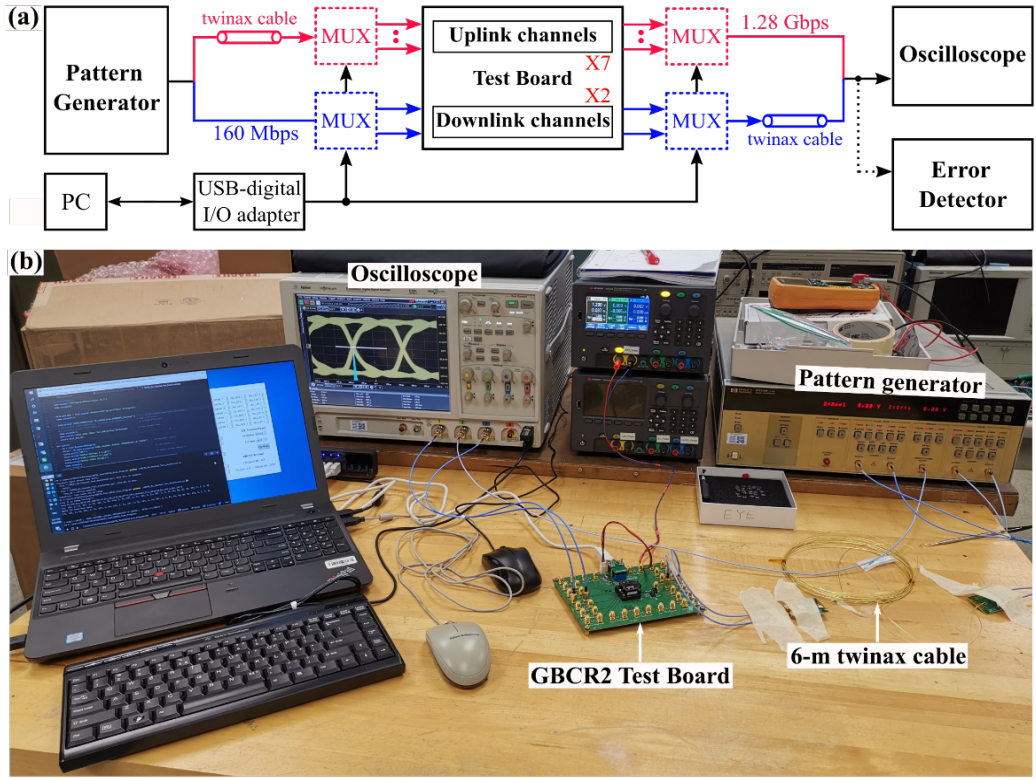

Figure 7. Block diagram (a) and photograph (b) of the test setup. The red line and blue line represent the uplink signal path and the downlink signal path, respectively.

### 3.2 Uplink channel test

As mentioned in Section 2, the equalization strength can be adjusted to compensate for the high-frequency signal loss and reduce the jitter by tuning the high-frequency and the medium-frequency source resistance (HFSR and MFSR) in the CTLEs. To find the optimal operation parameters, we scanned HFSR and MFSR and measured the eye diagram. To compute the jitter, the transition positions of the eye diagram around 0 mV are measured and plotted in a histogram. The jitter is calculated as the standard deviation of the histogram. Note that random jitter and deterministic jitter were not decomposed. The jitter heatmap plot is shown in figure 8. As can be seen from the figure, when the HFSR and the MFSR are out of the upper-left or the bottom-right region, the jitter is below 50 ps. The upper-left region represents that the equalization is too low, whereas the bottom-right region represents that the equalization is too high. The minimum jitter of 17.2 ps (RMS) is achieved when the HFSR and the MFSR are set to be 7 and 10, respectively, which are selected for all the following tests.

The phase shifter adjusts the clock phase for sampling the equalized data. To characterize the phase shifter, we measured the delay of the testing-purpose channel related to the input clock (both at rising edges) in different delay settings from 0 to 15. The measured relative delay versus the delay setting is shown in figure 9. It can be seen in the figure that the measured relative delay linearly increases with the increment of the delay setting. The slope of the linear trend line is about 47.9 ps, which is consistent with the theoretical value of 48.8 ps, 1/16 of the input clock period.

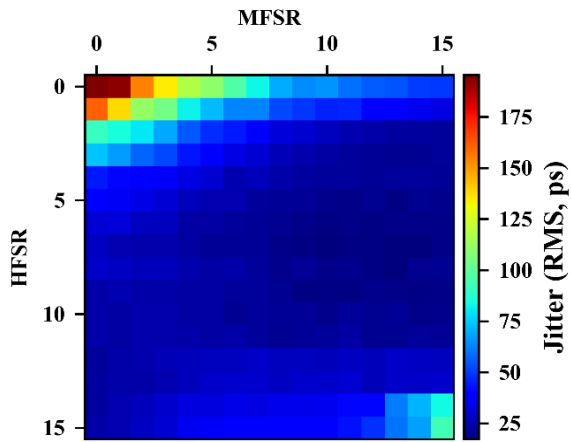 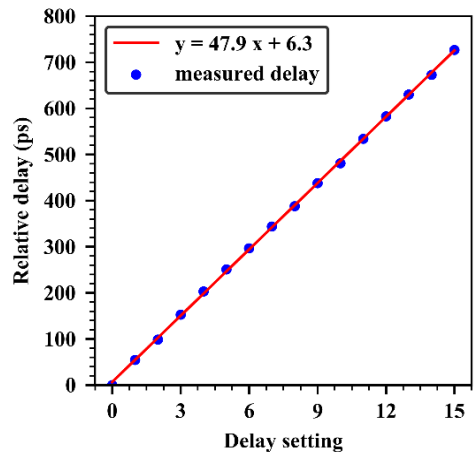

Figure 8. Jitter versus peaking parameters.  Figure 9. Measured relative delay versus delay setting.

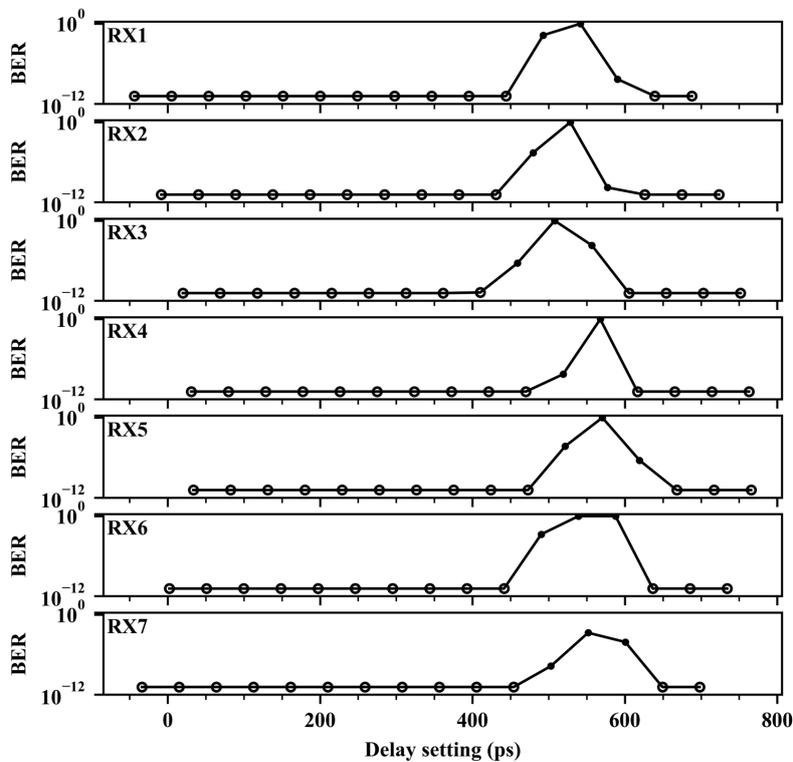

Figure 10. BER versus the corrected delay.

In the retiming mode, the sampling clock does not meet the setup/hold time requirements at certain delays and the data are not be sampled correctly. To find proper delay of data sampling, we studied the dependence of the BER on the delay. The delay ranged from 0 to 15 with the step of 1 (i.e., 48.8 ps). Each channel was measured for one minute at each delay with the optimal equalizer parameters. The measurement results are depicted in figure 10. On the test board, different uplink channels have different trace lengths, which are used to correct the relative delays. The x axis of the figure is the corrected delay. When no error is detected in one minute, $1.3 \times 10^{-11}$ is used to represent the upper limit of the BER. As can be seen in the figure, no error is observed in the whole range except at three delays. These three delays correspond to a transition window where the setup/hold requirements are not met. It should be noted that the transition windows of seven channels are aligned, meaning that a single delay setting can be applied to all uplink channels if the cable length is identical. For the uplink Channel 1, at the delay of 6, which is far away from the transition window, no error is observed for 15 hours. The corresponding BER is estimated to be $4.33 \times 10^{-14}$ with a confidence level of 95%.

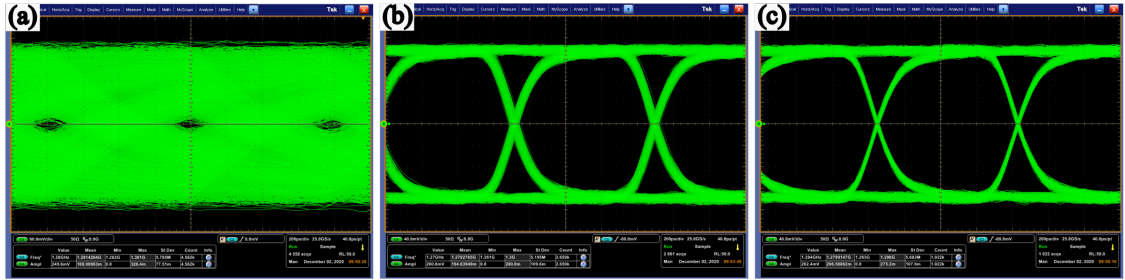

Figure 11. Eye diagrams before (a) and after recovering (non-retiming (b) and retiming mode (c)).

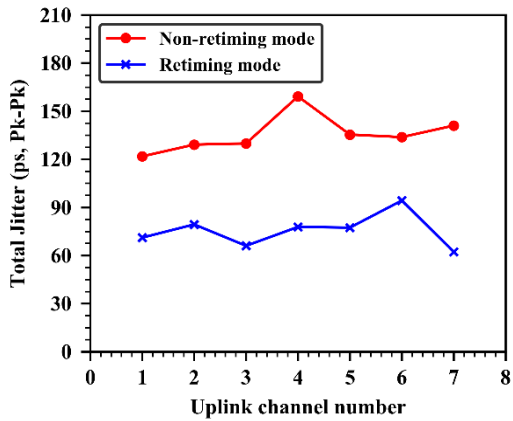

Figure 12. Total jitter of all uplink channels at the delay of 6.

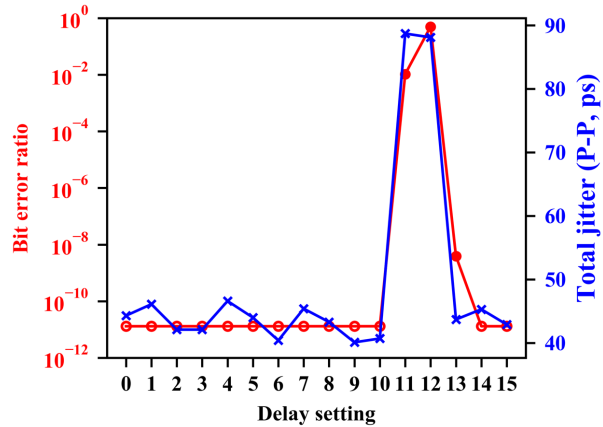

Figure 13. BER-Jitter correlation of uplink Channel 1.

Eye diagrams of uplink Channel 7 at different test points are shown in figure 11. When the signal from the pattern generator passes through the cable, the eye diagram is almost closed, as shown in figure 11(a). After being recovered by the GBCR2, the eye opens again, as shown in figure 11(b) in the non-retiming mode and figure 11(c) in the retiming mode. In the retiming mode, the delay is 6, far away from the transition window shown in figure 10. The jitter in the retiming mode is significantly smaller than in the non-retiming mode.

The jitter of all the uplink channels was analyzed with a jitter analysis program built in the oscilloscope. Random jitter and deterministic jitter were decomposed. In the retiming mode, the delay of all uplink channels was set to 6, far away from the transition window. The jitter of all uplink channels in the non-retiming mode and the retiming mode is shown in figure 12. As can be seen from figure 12, the Total Jitter (TJ, peak-peak) in the retiming mode is about half of that in the non-retiming mode. The TJ in both modes meets the specification (200 ps, peak-peak).

The correlation between the BER and the TJ of uplink Channel 1 at different delays is shown in figure 13. When the delays range from 11 to 13, the BER is higher than those in other delays. Similarly, the TJ at the delays of 11 or 12 is twice the values in the other delays. The transition window observed in the BER measurement is roughly consistent with the peak observed in the TJ measurement.

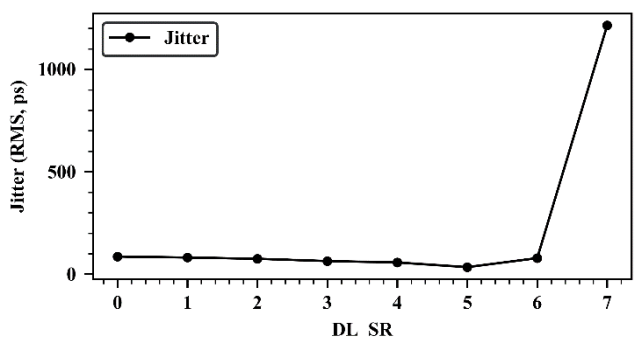

Figure 14. Dependence of jitter on the source resistance.

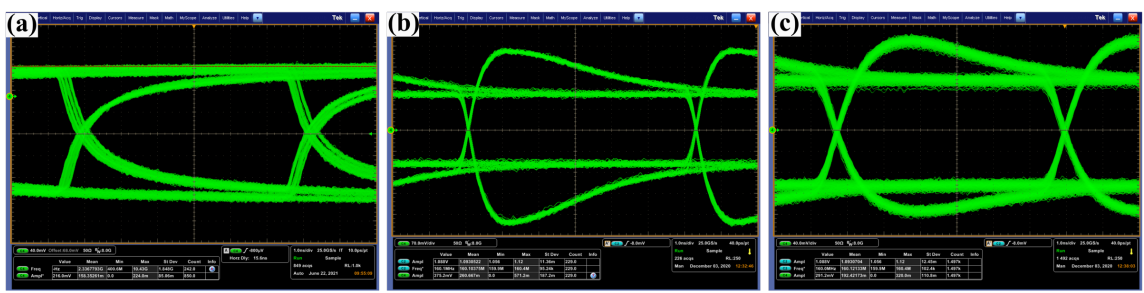

Figure 15. Eye diagrams after the cable without GBCR2 pre-emphasis (a), before (b) and after (c) the cable with GBCR2 pre-emphasis.

### 3.3 Downlink channel test

As mentioned in Section 2, the jitter of the downlink channels mainly depends on the source resistance when the pre-emphasizer is enabled. In order to obtain the optimal configuration, we swept the parameters and measured the jitter in the same way as in the uplink parameter scanning. The jitter versus source resistance (DL_SR) scanning is shown in figure 14. As can be seen in the figure, the jitter is less than 100 ps (RMS) except that the source resistance is the maximum. When the source resistance is 5, the jitter reaches the minimum value of 34.4 ps (RMS). The following tests for the downlink channels are performed under these optimal parameters.

The eye diagrams before and after the 6 m Twinax cable with the optimal parameters are shown in figure 15. Figure 15(a) shows the eye diagram when the pattern generator signal goes through the cable. As can be seen in figure 15(b), the overshoot in the eye diagram before the 6 m Twinax cable is obvious due to pre-emphasis. The TJ is 215 ps (peak-peak) before the 6 m Twinax cable. The eye diagram is still open after passing through the 6 m Twinax cable, as shown in figure 15(c). The TJ is 330.5 ps (peak-peak) after the 6 m Twinax cable. The total jitter after the cable is larger than the design specification of 200 ps.

The downlink jitter issue can be worked around. In the system test, when the lpGBT eTx pre-emphasis is combined with the GBCR2 pre-emphasis, the jitter of the RD53B output is less than when the GBCR2 is bypassed [16]. It seems that either the lpGBT eTx pre-emphasis or the GBCR2 pre-emphasis is strong enough. The combination of the lpGBT eTx and GBCR2 meets the system specification in terms of the BER. Due to the upgrade schedule constraint, the current GBCR2 downlink channel will be kept as the baseline design for the final production.

### 3.4 Power consumption

We studied the power consumption of all channels at three different supply voltages (1.08 V, 1.20

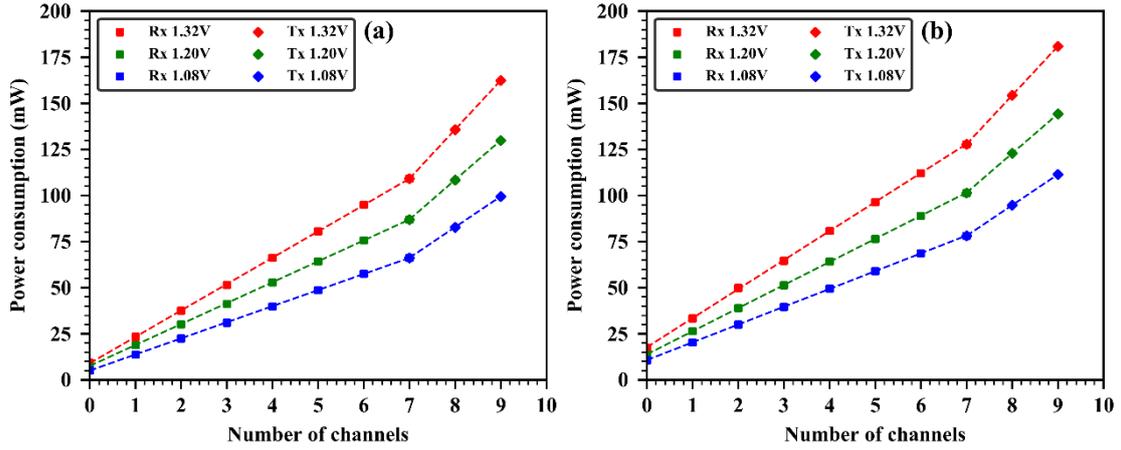

Figure 16. Power consumption in the non-retiming mode (a) and in the retiming mode (b).

V, and 1.32 V) in the non-retiming mode and the retiming mode, respectively. The power consumption versus the number of enabled channels is shown in figure 16. When the number of channels is 0, all the channels are disabled except the I$^2$C target and the phase shifter. The number of channels from 1 to 7 indicates the corresponding number of uplink channels are enabled. The number of channels from 8 to 9 denotes that all seven uplink channels are enabled and one or two downlink channels are activated. The fitting slopes for uplink channels and downlink channels are presented in Table 1. The power consumption per uplink channel in the retiming mode is 10% higher than that in the non-retiming mode. The power consumption and the supply current increase with the supply voltage from 1.08 V to 1.32 V. The total power consumption of all uplink channels is 109.1 mW in the non-retiming mode and 127.7 mW in the retiming mode, below the specification of 174 mW. The two downlink channels consume less than 53 mW.

Table 1. Power consumption of each channel.

| Voltage (V) | Uplink channel (fitting slope) | | Downlink channel (fitting slope) |
|---|---|---|---|
| | Non-retiming mode | Retiming mode | |
| 1.08 | 8.73 | 9.64 | 16.67 |
| 1.20 | 11.35 | 12.53 | 21.43 |
| 1.32 | 14.30 | 15.75 | 26.62 |

## 4. Quality control test

169 prototype chips are produced for Opto-Boxes tests. A Quality Control (QC) test procedure is proposed. The QC test is performed after the dies are packaged but before the packaged chips are assembled. The QC test aims at screening the components that meet the basic functional requirements and removing any components with issues caused by fabrication or packaging to reduce the cost of assembling failed chips.

### 4.1 Procedure of the QC test

In the QC test, eye diagrams of each uplink channel and downlink channel and the supply current are measured. The QC test procedures are as follows. First, each chip was placed in the socket on the test board, and the power supply was turned on with the current limited to double the nominal value (125 mA). If the supply current reached the maximum value, the chip was removed as a short failure. Second, preset values were written into the registers, and then the registers were read back via the I$^2$C interface. If the read values did not match the written values, this step was attempted again up to 10 times. The chip was removed as an I$^2$C failure if the read values were not consistent with the written values after ten I$^2$C operations. Third, the eye diagram of each uplink or downlink channel was collected, and the jitter was measured. The jitter should be less than 100 ps (RMS) both for uplink channels and downlink channels. Finally, each failed chip was double-checked up to twice to reduce the failure probability due to the contact issue with the socket.

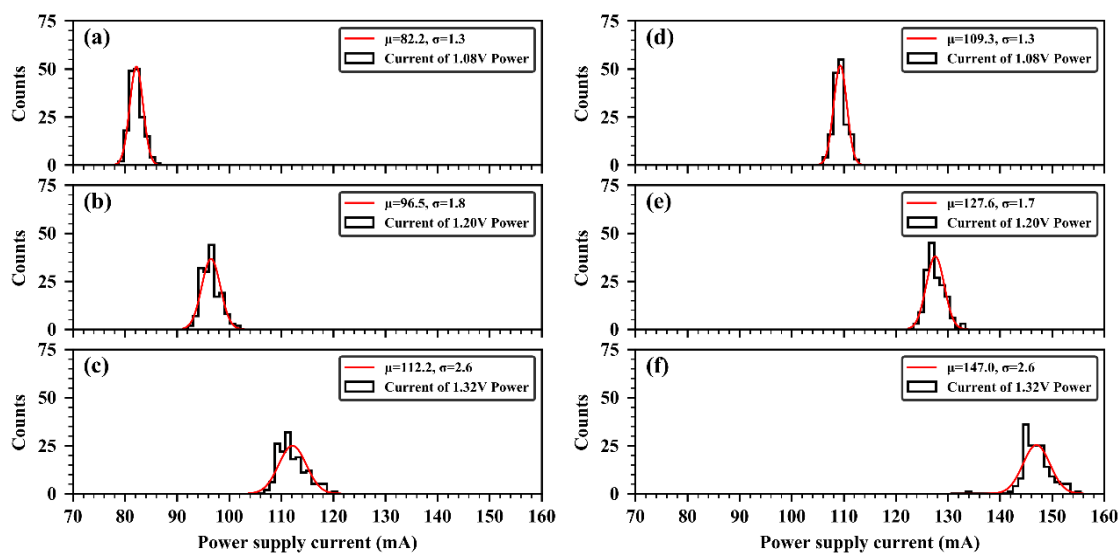

Figure 17. Supply current distribution for uplink channels (a)-(c) and downlink channels (d)-(f) at different supply voltages.

### 4.2 Test bench of the QC test

The block diagram of the test bench of the QC test is shown in figure 7(a). A RF multiplexer (Analog Devices, Part No. HMC321ALP4E) selected a specific uplink or downlink channel. A USB-to-digital-I/O adapter (LabJack, Part No. U3) controlled the RF multiplexer. The jitter of the eye diagram was measured in the same way as in the uplink parameter scanning.

### 4.3 Results of the QC test

Among the 169 chips under test, 164 chips were fully qualified. The other four chips each have one uplink or downlink channel eye diagram failure. The yield is about 97.0%.

Before testing all uplink channels and all downlink channels, we recorded the total supply current of all uplink and downlink channels at the supply voltages of 1.08 V, 1.20 V, and 1.32 V while providing the PRBS signal to a single channel. The distributions of the supply current in the retiming mode are shown in figure 17(a)-(c) for uplink channels and figure 17(d)-(f) for downlink channels. The power supply current obeys the Gaussian distribution, which is also displayed in the red fitting line. It can be seen that the supply current increases with the power supply voltage. The total power consumption of all uplink channels is 109.1 mW in the non-retiming mode and 127.7 mW in the retiming mode, below the specification of 174 mW. The two downlink channels consume less than 53 mW.

The jitter distributions are plotted in figure 18(a)-(c) for all uplink channels in the non-retiming mode and figure 18(d)-(f) for all downlink channels at the supply voltages of 1.08 V, 1.2 V, and 1.32 V. Since GBCR2 has seven uplink channels and two downlink channels, the counts of the distribution for uplink channels are nearly 3.5 times of the downlink channels. When the supply voltage ranges from 1.08 V to 1.32 V, the jitter does not significantly change.

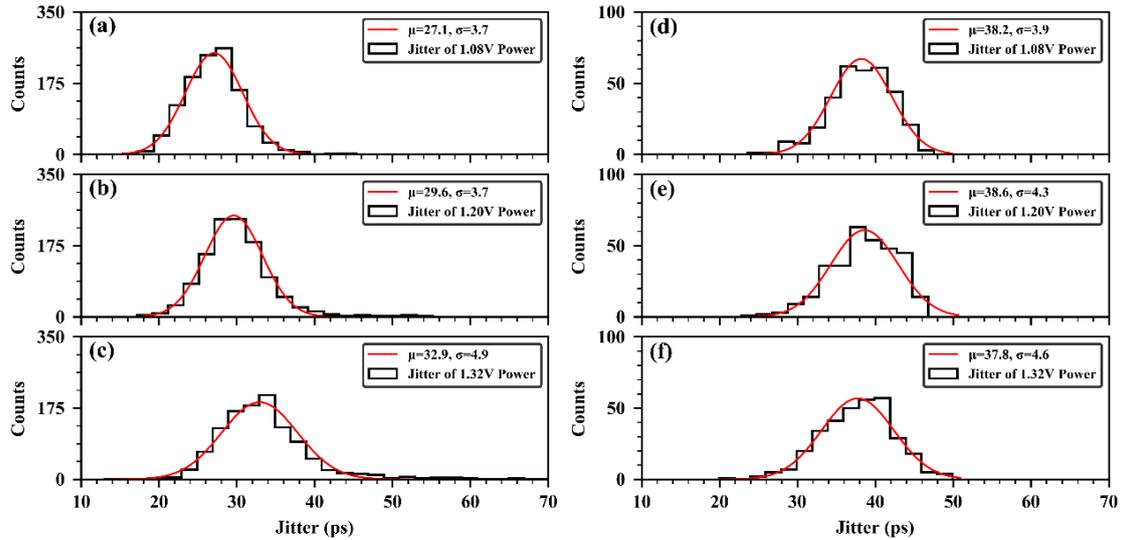

Figure 18. Jitter distribution for uplink channels (a)-(c) and downlink channels (d)-(f).

## 5. Conclusion

We have successfully characterized GBCR2 for the ATLAS ITk pixel detector Phase-II upgrade. For the uplink channels, the total jitter of the output signal is 129.1 ps (peak-peak) in the non-retiming logic and 79.3 ps (peak-peak) in the retiming mode, respectively. The total power consumption of all uplink channels is 109.1 mW in the non-retiming mode and 127.7 mW in the retiming mode, below the specification of 174 mW. The two downlink channels consume less than 53 mW. The test results demonstrate that GBCR2 works as expected. We have proposed a QC test procedure. In a small batch of 169 prototype chips, 164 prototypes are qualified. The yield is 97.0%. 125 chips have been assembled in optical boxes.

## Acknowledgments

This work was supported by the US-ATLAS phase-2 upgrade grant administrated by the US-ATLAS phase-2 upgrade project office and the Office of High Energy Physics of the U.S. Department of Energy under contract DE-AC02-05CH11231. We are grateful to Dr. Maurice Garcia-Sciveres and Ms. Veronica Wallangen for providing the cable model data and Dr. Dong Su, Dr. Andrew Young, and Dr. Zhijun Xu for fruitful discussion. The authors would also like to thank Dr. Szymon Kulis and Dr. Paulo Moreira for contributing to the design of the I$^2$C target design block.